\documentclass[a4paper]{jpconf}

\usepackage{iopams}  
\usepackage{graphicx}

\begin{document}

\title{Conductance Anomalies in Quantum Point Contacts
and One Dimensional 
Wires}

\author{Mukunda P. Das$^1$ and Frederick Green$^2$}

\address{$^1$ Department of Theoretical Physics,
RSPE, The Australian National University, Canberra, ACT 2601, Australia.}
\address{$^2$ School of Physics, The University of New South Wales,
Sydney, NSW 2052, Australia.}

\begin{abstract}
Over the last decade, interest in one-dimensional charge transport has
progressed from the seminal discovery of Landauer quantization of
conductance, as a function of carrier density, to finer-scale phenomena at
the onset of quantization. This has come to be called the ``0.7 anomaly'',
rather connoting a theoretical mystery of some profundity and universality,
which remains open to date. Its somewhat imaginative appellation may tend to
mislead, since the anomaly manifests itself over a range of conductance
values: anywhere between 0.25 to 0.95 Landauer quanta. In this paper we offer
a critique of the 0.7 anomaly and discuss the extent to which it represents a
deep question of physics.
\end{abstract}

\section{Introduction}

The live topic of the ``0.7 anomaly'' in quantized quasi-one dimensional (1D)
conductance has occupied both experimentalists and theorists already for some
time, and continues to do so. The effect, or at least effects very like it,
has been studied over a wide spectrum of mesoscopic material structures and
fabrication techniques. So far, no conclusive theoretical reason has been
adduced to explain the origin of the anomaly, any more than the universality
implicit in its title.

Electronic transport in quasi-one dimensional systems is essentially studied
in quantum-point contact (QPC) and quantum-wire structures. The quantized
nature of their conductance, known as Landauer quantization, is a genuinely
generic 1D phenomenon, first reported in QPCs of high-electron-mobility
devices in GaAs/GaAlAs heterostructures. As a function of carrier density
(modulate by a transverse gate potential), distinct steps in the differential
conductance (G is normalized to the quantum value $G_0 = 2e^2/h$)
were observed \cite{1,2}.

Various explanations for the quantized Landauer steps have been given over the
years. Those comprise the simpler Landauer-B\"uttiker method, the more
rigorous Kubo linear-response formula and finally the so-called
nonequilibrium-Green-function technique. We cite two of our relevant
papers \cite{3,4} presenting an appraisal of quantized-conductance physics.

From 1988 onwards, a much wider set of different 1D systems, from Si
metal-oxide field-effect transistors \cite{5}, other III-V heterostructures \cite{6},
and constricted graphenes \cite{7} etc. have been shown to exhibit these quantized
steps. However, apart from the quantized steps now well established as
universal in nature (occurring at integral values $n= 1, 2, 3$ etc. for
$G/G_0 = n$), there also seem to be anomalies observed in many of the above
systems. These involve peaks and shoulders at non-integral values of G,
located on the threshold regions of gate voltage, at the onset of a new
quantized ``Landauer'' step.
Generally there are two types of such anomalies.

\begin{itemize}
\item
Thomas {\em et al} \cite{8} originally identified an anomalous structure in
low-temperature conductance below the first quantized step in GaAs QPCs at
about $0.7G_0$.  It has been argued that this perceived anomaly is an\
intrinsic
property caused by some many-body effect independent of the details of
the material system.

The authors of Ref. 8 argued that, since the effect appears to occur in the
majority of one- dimensional conduction channels, it may be universal; this
is what the majority of the literature, subsequent to Ref. 8, likewise
suggests. Although the anomaly has been discussed abundantly ever since,
it should be pointed out that many such features have now been subsumed into
the 0.7 rubric even when it is abundantly clear that anomalies occur in all
kinds of temperature ranges and at nonzero magnetic fields.

\item
 A further, and very different, anomalous peak has been identified in the
nonlinear transport regime at low temperatures -- known as the zero-bias
anomaly (ZBA) in the differential conductance -- as a function of source-drain
voltage along the channel. Unlike the range of so-called 0.7 features, this
anomaly is well understood and we shall not pursue it here.
\end{itemize} 

So far, published explanations for the physics of the 0.7 anomalies follow
two principal ideas: (1) the assumption of spontaneous spin polarization
(SSP) of the carrier population, and (2) the presence of a many-body state
induced by Kondo physics. 

SSP is a semi-phenomenological description \cite{9,10}. Spin-polarized
density-functional calculations by Berggren and coworkers \cite{9,10} have shown
that exchange interactions can induce a large sub-band splitting whenever the
Fermi energy passes through the sub-band threshold energies. Non-monotonic
variations in the densities of states, as the gate potential varies, leads to
different degrees of spin polarization and are liable to produce conductance
anomalies.

In the current literature both 0.7 and ZBA features have been explained
\cite{11,12} by either: amplification of interaction effects when a smeared
van-Hove singularity of the local density of states (at the bottom of the
lowest 1D sub-band) crosses the chemical potential; or by an emergent
localized state contributing to the formation of a collective (many-body)
state arising from the Kondo effect in electron transport through the
localized state. Another work, Ref. 13, develops a similar idea based on
Kondo physics. Here we note that the latter mechanisms both run counter to
the alternative models assuming spontaneous spin polarization, and shall
return to these questions below.

In the next Section we give a pedagogic picture of the principal Landauer
conductance steps for a long quantum wire or quantum point contact. In Sec. 3
we discuss various non-integral features of conductance as a function of
split-gate voltage, for a variety of 1D systems. We present a brief review
in Sec. 4, with full reference to the current literature as to how one
attempts to understand the anomalous features. In Sec. 5 we conclude our
study with a critical discussion study of the presumed universality of the
so-called conductance anomalies, arguing that they do not meet any reasonable
measure of what a truly generic phenomenon should be. In doing so we refute
claims \cite{14,15} that these anomalies are fundamentally intrinsic to 1D
electronic transport. They are, by contrast, material specific not at all
``universal''.

\section{Quantum wires and QPCs: conductance}

Quantum wires and QPCs are two examples of real quasi-one-dimensional
electronic  systems \cite{16}. They are fabricated by a variety of established
techniques: cleaved-edge overgrowth (CEO) for the former
\cite{Z}, and split-gate geometry for the latter.
To interpret the observed conductance behaviour of
these devices we go to Landauer's model.

The schematic of Figure 1 shows how a device is connected to
its external current source by two large reservoirs, which
are assumed to acquire two unequal chemical potentials when biased, being
mutually offset by the applied electrostatic energy $eV$.
The 1D system has a long (ideally infinite) length supporting several
electronic conduction sub-bands, depending on the width of
constriction or, alternatively, the strength of side-gate potential
which modulates the 1D carrier confinement. A top gate normally
supplies a transverse bias potential that modulates the 1D carrier
density within the channel.


\centerline{
\includegraphics[width=8truecm]{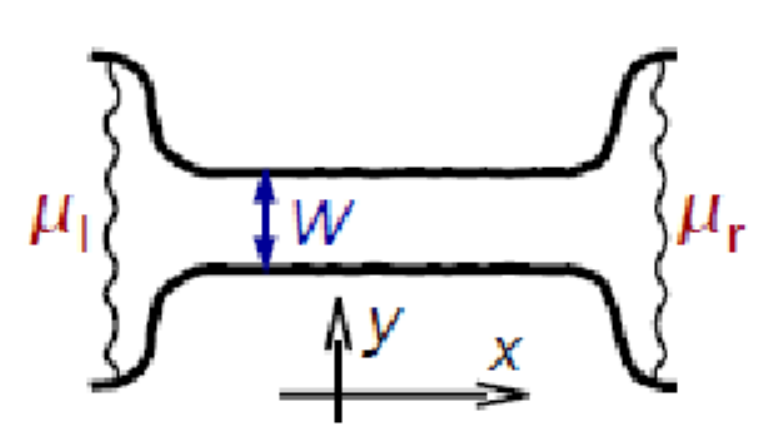}
}
\vskip 0.25 truecm 
{\small {\bf FIG. 1}.
Schematic view of an ideal quantum wire or quantum point contact.
The wire width $W$ is small relative to the effective length of the
channel. Current is presumed to flow if the left-hand reservoir
chemical potential $\mu_l$ is raised above the right-hand value $\mu_r$.
The offset $\mu_l - \mu_r$ is equated with the externally measured
electrostatic energy $eV$ per carrier. In each supported sub-band,
electrons may occupy current-carrying states with wavevector $k$
along the axis of the channel.}
\vskip 0.25 truecm 

The channel in a ``perfect'' device is long and uniform and
perfectly transmissive.
The net electron current per channel per spin is assumed to be
the difference in integrated flux $-ev_k$
between unidirectional ``right-movers''
with distribution $f_r = f(\epsilon_n(k) - \epsilon_F - \mu_r)$
and ``left-movers'', $f_l = f(\epsilon_n(k) - \epsilon_F - \mu_l)$.
Here $\epsilon_n(k) = \epsilon_n + \epsilon(k)$ is the sub-band energy
with $\epsilon_n$ the threshold and $\mu_l, \mu_r$ are the left and right
reservoir electro-chemical potential respectively.
With Pauli blocking of the complementary final states, the result is

\begin{eqnarray}
J_n
&=&
\int^{\infty}_0 {dk\over 2\pi} (-ev_k)
{\Bigl( f_r(1 - f_l) - f_l(1 - f_r) \Bigr)}
\cr
&=&
-{e\over 2\pi\hbar} \int^{\infty}_0 d\epsilon \delta\mu
{\partial \over \partial \epsilon} f(\epsilon + \epsilon_n - \epsilon_F)
\cr
&=&
{e^2\over h} f(\epsilon_n - \epsilon_F) V
\label{eq1}
\end{eqnarray}

\noindent
where one also assumes the offset of reservoir electro-chemical
potentials $\delta\mu = \mu_l - \mu_r$ to be
the electrostatic energy per carrier, $eV$,
gained from the externally applied potential difference.

In the zero-temperature limit $f(\epsilon_n - \epsilon_F)$
is zero if the Fermi level falls below the sub-band threshold,
and unity if the level is above threshold and the sub-band is populated.
With spin degeneracy and with $N$ sub-bands populated,
one then obtains the total conductance as  $G = (2e^2/h) N \equiv G0 N$.

The resistance of this ideal system is $G^{-1}$, which is finite, independent
of length, and indeed universal. When the chemical potential crosses the
bottom of each sub-band, the conductance ratio $G/G0$  steps up from 1 to 2,
and so on. These are the well known quantized conductance steps (see Fig.1). 

The flux $-ev_k$ in the integrand on the right-hand side of Eq. (\ref{eq1})
above is further modified in the presence of quantum coherent
scattering (unitary, hence nondissipative), accounting for partial
reflection of incoming states and lowering the net transmission
below ideal. The conductance in a single
channel typically becomes $G = G_0{\cal T}$, where $0 \leq {\cal T} \leq 1$.
When the transmission factor ${\cal T}$ is non-ideal, the conductance
will be seen to undershoot its full integer value.


\centerline{
\includegraphics[width=6.0truecm]{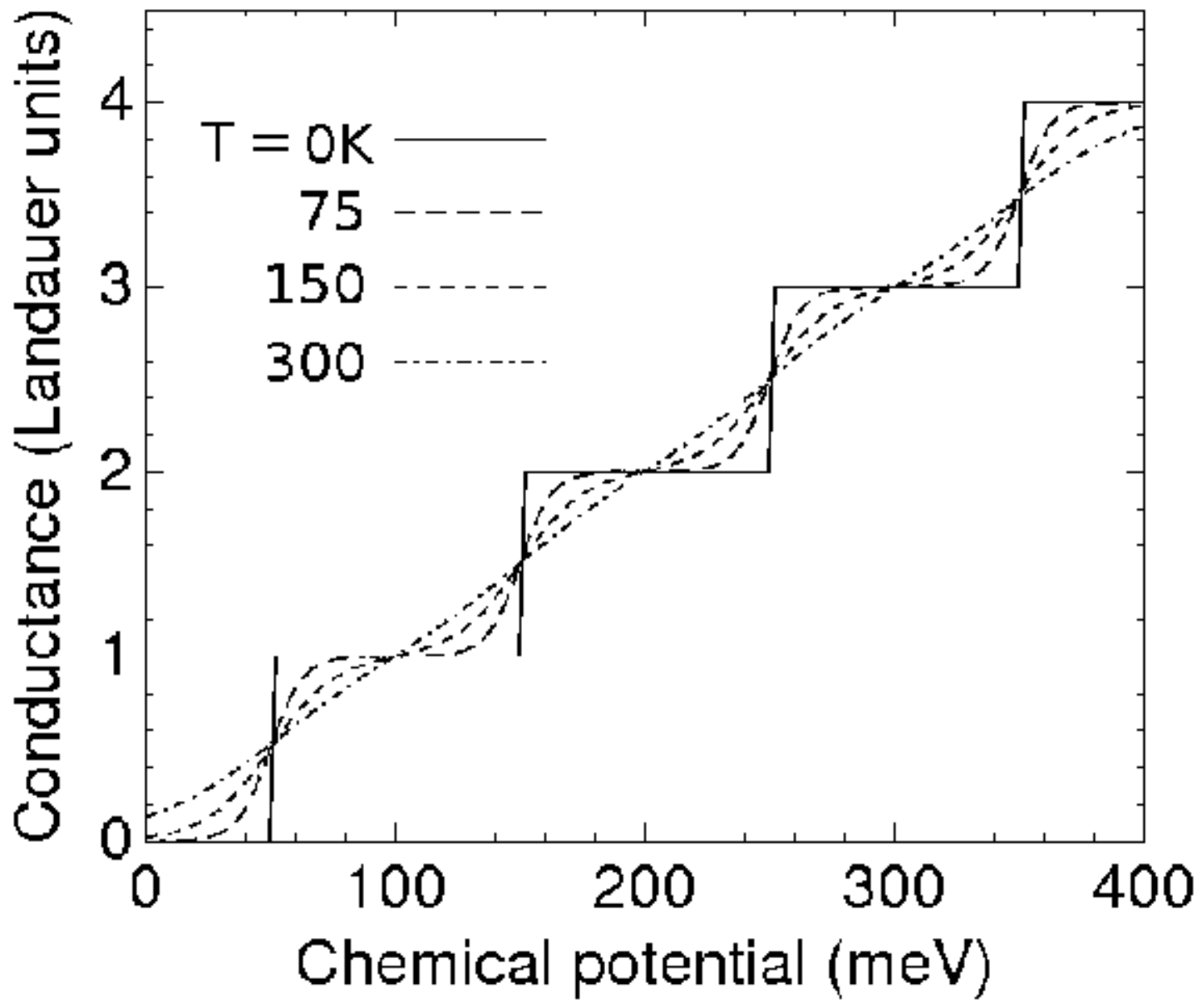}
}
%
{\small {\bf FIG. 2}.
Ideal Conductance $G$ of a QPC as a function of chemical potential for
various temperatures; see Equation (\ref{eq1}) in the text.
The transverse confinement is modelled as a
harmonic potential with quantized energy $\hbar\omega=0.1 eV)$.
The characteristic Landauer steps in $G$
have been confirmed in many different quantum wires and QPCs. A
major issue remains unresolved within the quantum-coherent and strictly
one-particle theory of the steps: What is the mode of
its physical dissipation (Joule heating)?
Since the Landauer theory admits only coherent elastic scattering,
any physical mechanism for actual dissipation is missing.}
\vskip 0.20 truecm 

The commonly accepted phenomenology of $G$ is known as the Landauer
(sometimes, Landauer-B\"uttiker) model and is
overwhelmingly applied in most analyses of real measurements. It
asserts \cite{16} that the dissipation takes place exclusively the leads. If
so, however, a further question becomes inevitable at ideal
transmission $T = 1$: What relevance can the internal physics of a coherent
quantum channel have, in that case, for the (universal) Joule-heating
formula $P = IV = GV^2$? 

In Fig. 3 we show the quantized conductance steps as a function of chemical
potential. Differently from the Landauer model, these are calculated
within a standard quantum-kinetic description of a 1D wire,
in which inelastic dissipation enters naturally and concurrently with
elastic scattering. The model is an application of the canonical Kubo
formula for transport.

\vskip 0.5 truecm 
\centerline{
\includegraphics[width=6.5truecm]{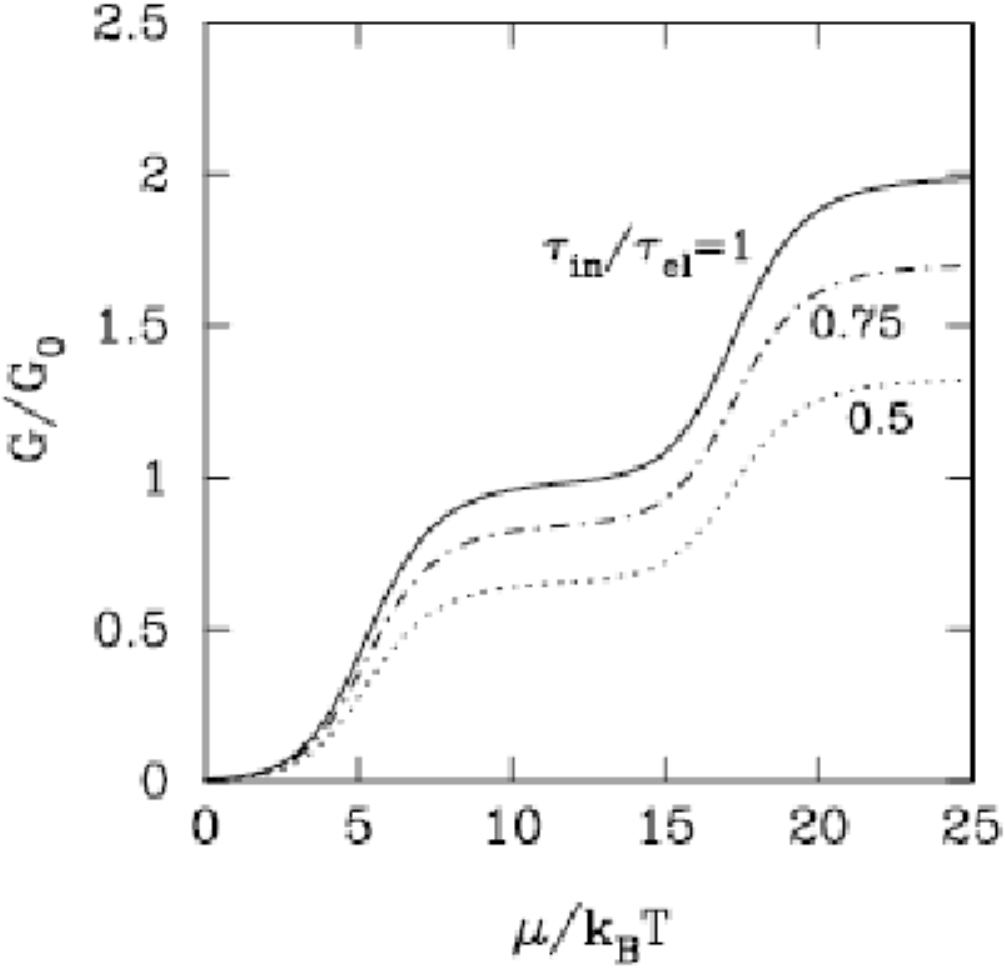}
}
\vskip 0.20 truecm 
{\small {\bf FIG. 3}.
Ideal and non-ideal behaviour of quantized conductance $G$ in a 1D confined
channel as a function of carrier chemical potential. The figure
is taken from Ref. \cite{W}. In contrast with the coherent
transmission model for
$G$, full quantitative allowance is made for dissipative scattering even
at ``ideal'' conduction, where the mean free paths for elastic and
inelastic scattering are perfectly matched to the operational length
of the channel. When these lengths are mismatched, the conductance becomes
non-ideal.
\vskip 0.20 truecm 


\section{Anomalies at non-integer values of $G$}

In the introduction we highlighted the existence of anomalies in the value of
conductance at specific values of gate voltages and thus chemical potentials
in the channel.  The following (partial) list gives an indication of the
non-integral ``anomalous'' values of G in a range of device situations.
\vskip 0.20cm

\begin{tabular}{|c|c|l|}
\hline
Structural type	& Anomaly at/around & ~~~~~~~~~ Comments \\
\hline
                &     & Kink disappears	when T\\
GaAs/GaAlAs QPC	& 0.7 & is lowered; when magnetic\\
\cite{8}        &     & field is applied the anomaly\\
                &     & moves from 0.7 to 0.5. \\
                &     &\\
\hline
            & 0.4 &   at  ~0.0mV source-drain voltage\\
Si-SiGe QPC & 0.6 & ~$''$ ~2.0mV   $\! ~~~~~~~'' ~~~~~~~~~~~~''$\\
von Pock {\em et al} \cite{5}
               & 0.2 & ~$''$ 10.0mV   $\! ~~~~~~'' ~~~~~~~~~~~~''$\\
            &     &\\
\hline
         &             & in range of $T$ from 5K to 24.8K.\\
InAs QPC & 0.15 to 0.7 & Anomalous structures are \\
Lehmann {\em et al} \cite{6}
         &             & 0.15, 0.5, 0.7 etc. for\\
         &             & mag. fields from 0 to 7 Tesla\\
         &             &\\
\hline
GaAs-GaAlAs  & 0.66 to 0.8  & at 1K; structures vanish at\\
quantum wire &              & higher $T$\\
dePicciotto {\em et al} \cite{Y}
             &              &\\
\hline
\end{tabular}
\vskip 0.20cm

{\small {\bf TABLE 1} Summary of measured
values and behaviours for the ``0.7'' anomaly, over a variety
of materials and structures.}
\vskip 0.25cm

The actual locations and magnitudes of these anomalies are evidently many and
varied. The point to be emphasised here is  that 0.7 G0 is hardly a universal
fixture in the classic sense of Landauer quantization. Nevertheless, the 0.7
anomaly, which in reality is highly contingent phenomenon, to date retains a
mystique of universality in a now considerable body of literature.


More recently Brun {\em et al} \cite{X}
have demonstrated the existence of a 0.7$G_0$
anomaly in a 1D QPC (fabricated on a GaAs/AlGaAs heterojunction structure)
together with a zero-bias anomaly. This appears consistent with a Kondo
mechanism, as we discuss below. For details, see the Caption of Fig. 4.
\vskip 0.25 truecm 

\centerline{
\includegraphics[width=7truecm]{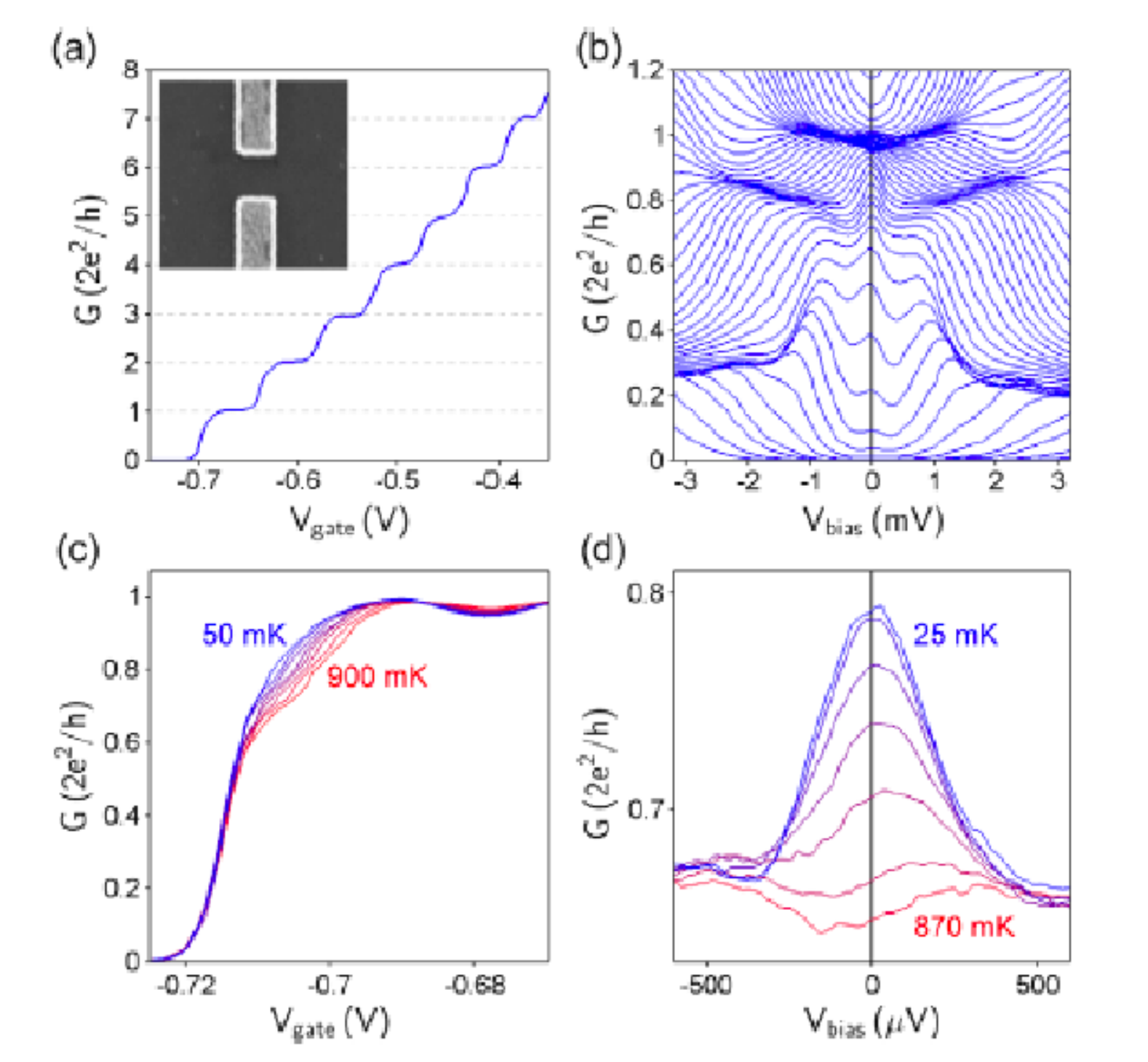}
}
\vskip 0.25 truecm 

{\small {\bf FIG. 4}.
QPC conductance versus gate voltage at 30mK.
Inset: image of the metallic split gate. (b) Differential conductance
versus source-drain bias at 25mK and different gate voltages.
(c) Temperature dependence of the 0.7 anomaly from 50 to 900mK.
(d) Temperature dependence of the zero-bias anomaly from 25 to 870mK.
Figure from Ref.\cite{X}.
}

\vskip 0.25 truecm

\section{Theoretical arguments}

Attempts to understand the anomalous conductance features at non-integral
values are already numerous and include, in the main: spontaneous spin
polarization; Kondo and van Hove types of singularity at/around the Fermi
level; Majorana modes; Wigner crystallization etc. There are more besides,
and we discuss here the better-known theoretical models popularised in the
existing literature.

\subsection{Spontaneous spin polarization}

Consider a two-dimensional electron system on the interface of a GaAs/AlGaAs
heterojunction material \cite{10}. Motion in the perpendicular direction to
the heterojunction interface (say in the $z$-direction) is strongly
inhibited by transverse quantum-well confinement, whose bound-state energies
are large. An additional, somewhat weaker quantum-well confinement
is imposed in the $y$-direction by constriction either from side gates
or by channel etching. In a harmonic-potential approximation the
corresponding sub-band levels in the $y$-direction would be
$E_n = \hbar (n +1/2)$ for $n = 0,1,2$ etc.

The conductive channel now extends indefinitely in the $x$- direction,
leading to a quasi-continous band of
extended, and thus current-carrying, states.
Consider a specific sub-band in the heterojunction.
The sub-band states' effective Schr\"odinger equation, within the Kohn-Sham
local-spin-density approximation (LSDA), is

\begin{eqnarray}
{\left[
{{p^2_x + p^2_y}\over 2m^*} + V_{\rm conf}(y) + V^{\sigma}_{\rm exch}(x,y) 
\right]} \varphi^{\sigma}(x,y)
=&&
E^{\sigma} \varphi^{\sigma}(x,y)
\label{eq2}
\end{eqnarray}

\noindent
where $\sigma$ refers to the carriers' spin and $m^*$
is their effective mass. Here also $V_{\rm conf}(x,y)$ represents the
lateral confinement potential (as well as the carriers' mean-field
(Hubbard) electrostatic potential). In its bare form, the confinement
part is taken to be

\[
V_{\rm conf}(x,y) \sim
{m^*\over 2} (\omega^2_y y^2 - \omega^2_x x^2) + V_0. 
\}\]

\noindent
The exchange potential $V^{\sigma}_{\rm exch}(x,y)$, on the other hand,
is intrinsic to interacting fermionic carriers. Exchange has a
particular form within the LSDA theory, and interested readers may
find the details in Ref. 10.

Numerical Kohn-Sham LSDA calculations based on Eq. (\ref{eq2})
were carried out in a
series of papers by Berggren and co-workers \cite{9,10}. The
conductance G as a function of the height of the height $V_0$
of the saddle point in the bare confinement potential
is calculated (Fig. 4 of Ref. 10). In the noninteracting Landauer model the
ratio $G/G_0$ goes through a step of unity in a certain range of values of
$V_0$; but in the spin-polarised exchange model, that ratio acquires
a Zeeman-splitting substructure at 0.5.  

Several groups \cite{17}--\cite{19}
have suggested that electron pairs are coupled in
singlet and triplet configurations, experiencing different transmission
barriers. Within a stochastic approach to the occupancy, the ratio of triplet
to singlet resonances will be 3:1. This suggests that there are two
structures in the plateau-to-plateau transition: at $G = 0.75G_0$
and $G = 0.25G_0$.

The calculations just cited proceed within the exchange-only LSDA theory. 
Later Berggren et al. \cite{20} included the short-range correlation potential
(that is, beyond Hubbard) by using a
parametrised correlation part from Monte Carlo data of Ceperley and Tanatar
\cite{21}. Although the exchange contribution dominates, short-range Coulomb
correlations have an important role in determining the spin-split QPC
potential. As a result, the calculated conductance as a function of
gate voltage shows two anomalies in the first plateau-to-plateau
transition, one at $0.4G_0$ for a lower gate potential and another at
$0.75G_0$ at a somewhat higher gate potential.

We note here that the spin-split picture here applies for free carriers
in a perfect 1D channel. This is in contrast with the picture
where the presence of an impurity is pivotal. We now discuss this theory.

\subsection{Kondo physics in a QPC}

The Kondo effect arises from the interaction of a single magnetic-impurity
atom, via its localised spin, with the surrounding conduction electrons of a
nonmagnetic metal. This interaction induces a drop of electrical resistance
at a particular temperature, which rises again as the temperature is further
lowered. Kondo's theory explained this anomalous behaviour in resistance for a
large number of nonmagnetic systems having dilute magnetic impurities. The
key idea is the presence of the localised spin interacting in a conducting
continuum.

Recently the Kondo effect has also been observed in quantum dot systems. A
quantum dot with at least one unpaired electron acts as a magnetic impurity
and the conduction electrons can scatter off the dot. This is analogous to
the Kondo effect outlined above. In a parallel development, the possibility
of a Kondo-like mechanism has been argued to have relevance to the so-called
conductance anomaly as well \cite{11}--\cite{13}, \cite{X}, \cite{22, 23}.
For a QPC this requires the
underlying assumption that the system possesses a {\em localised} Kondo-like
unpaired spin.

We examine the putative Kondo connection with the 0.7 anomaly. Its physics
differs markedly from the previous spin-based idea that the anomaly
is qualitatively explained by spontaneous spin polarization,
not requiring an embedded, magnetically active  localised impurity.
A number of papers has appeared in favour of a Kondo connection
with the 0.7 anomaly.

Cronenwett {\em et al} \cite{24} performed conductance measurements
in a GaAs/AlGaAs
QPC. Apart from the regular conductance quantization,
they also found an anomaly
around $0.7G_0$. Their argument goes against the two spin channels' being
simultaneously occupied. Since the 0.7 structure is observed to disappear
with lower temperatures they ascribed this behaviour to the Kondo effect (in
addition to the presence of a zero-bias conductance peak). Further
investigations suggested that the 0.7 anomaly and ZBA share certain
similarities with the Kondo effect in quantum dots \cite{12,13}. 

The analysis based on the Kondo approach is outlined here, with the model
Hamiltonian \cite{25}

\begin{eqnarray}
H
=&&
\sum_{\sigma; k \in L, R}
\epsilon_{k\sigma} {\bf c}_{k\sigma}^{\dagger} {\bf c}_{k\sigma}
+ \sum_{\sigma } \epsilon_{\sigma} {\bf d}_{\sigma}^{\dagger} {\bf
  d}_{\sigma}
+ U {\bf n}_{\uparrow} {\bf n}_{\downarrow}
\cr
\cr
&&
+ \sum_{\sigma; k \in L, R} {\Bigl[
V^{(1)}_{k\sigma}
(1-{\bf n}_{\overline\sigma}){\bf c}_{k\sigma}^{\dagger}{\bf d}_{\sigma}
+ V^{(2)}_{k\sigma}
{\bf n}_{\overline\sigma}{\bf c}_{k\sigma}^{\dagger}{\bf d}_{\sigma}
+ {\rm H.c.} \Bigr]}
\label{eq3}
\end{eqnarray}

\noindent
Equation (\ref{eq3})
is a canonical Anderson Hamiltonian with $\epsilon_{k\sigma}$
and $\epsilon_{\sigma}$ the energies of
conduction and localised electrons respectively. $U$
is the on-site Coulomb interaction and the $V$s control
the hybridization of conduction and localised electrons.
The subsidiary notations are detailed in Ref. 23.

By performing a Schrieffer-Wolff transformation on this Hamiltonian \cite{23}
one obtains a Kondo Hamiltonian with coupling (to second order)

\[
J^{(i)}_{kk'; \sigma\sigma'}
= {(-1)^{i+1}\over 4}
{\left[
{V^{(i)}_{k\sigma}{V^*}^{(i)}_{k'\sigma'}\over
{\epsilon_{k\sigma} - \epsilon^{(i)}_{\sigma}}}
+ {V^{(i)}_{k\sigma}{V^*}^{(i)}_{k'\sigma'}\over
{\epsilon_{k'\sigma'} - \epsilon^{(i)}_{\sigma'}}}
\right]},
\]

\noindent
which is a reminder that the Kondo model is, in fact,
a special case of the more general Anderson impurity model.

The associated expression for the 1D Kondo conductance
can be derived from the tunnelling theory of Appelbaum \cite{25}:

\begin{eqnarray}
G_2
=&&
{4\pi e^2\over \hbar} \rho_L(\epsilon_F)\rho_P(\epsilon_F)
{\Biggl\{ (J^{(-)}_{LR})^2 + (J^{(+)}_{LR})^2
\Biggr.}
\cr
\cr
&& \times
{\Biggl.
{\left[ 3 + 2{\langle M \rangle}
{\left( {\rm tanh}{{\Delta + eV}\over 2k_{\rm B}T}
      + {\rm tanh}{{\Delta - eV}\over 2k_{\rm B}T} \right)}
\right]}
\Biggr\}}.
\label{eq4}
\end{eqnarray}

\noindent
Again, the detailed notations are detailed in Ref. 23. By adjusting
the several free parameters of the theory, namely
$U, V^{(1)}, V^{(2)},$ etc. (see Fig. 3 of \cite{23})
one may reproduce theoretically a dip in $G$ as a function of gate
voltage (modulating the channel's chemical potential $\epsilon_F$ )
in the range 0.5 -- 0.75$G_0$. 
Occurrence of the anomaly in the conductance, particularly at
0.7, depends overwhelmingly on the choice of a variety of
contingent phenomenological parameters, as we point out above. 

We recapitulate. The key idea of the Kondo picture for a QPC,
and the relevance of Eq. (\ref{eq4}),
requires the presence of a localised impurity. However, this
precondition has been put into question, particularly for clean QPCs.
More recent works following this type of argument have appeared
\cite{11,12,13,X}.

\subsection{Van Hove singularity}

Taking another, perhaps simple-minded line, if one looks at the
density of states (DOS) of a QPC, the DOS, albeit partly smeared,
has some peak structures reminiscent of van Hove singularities.
The expression of conductance G, in all of the extant
theories, contains the density of states as its leading component.
If the Fermi level occurs in the high-density-of-states regime,
we may expect some structures to appear, corresponding to the
anomaly one is trying to understand.

In the case of the Kondo effect, such high DOSs also occur at the Fermi
energy. We note here that the van Hove singularity has no connection with
impurity states; they are generic to regular lattice structures.
Thus, even in a one-dimensional independent-electron model, singularities
will occur at the band edges.

\subsection{Other theoretical suggestions}

In hybrid semiconducting superconductor-nanowire devices involving a QPC,
tunnelling features are observed to manifest a ZBA along with other
conductance features, such as the anomaly around $G = 0.7G_0$.
With the current interest in unusual topological states of
matter and non-abelian quasi-particle statistics,
in this context one looks for an interpretation bringing in
Majorana zero modes \cite{26}. Nevertheless, to date these experiments
cannot clearly resolve differences between Kondo and Majorana physics.

A promising and attractive prospect is to study more deeply the explicit
electron correlations in QPCs, since no noninteracting- carrier model
is able to produce the conductance anomalies. At low electron
density, spontaneous electron localization may occur below the first plateau
\cite{27}. This resembles the occurrence of one-dimensional Wigner crystallization
\cite{28}. The picture of electron localization may, in any case,
be argued in support of the Kondo picture of 0.7-like anomalies.

\section{Summary and Conclusions}

We have presented a brief overview of the conductance anomaly,
particularly at non-integral values of conductance at or just below
the first Landauer-step transition. We began by examining
quasi-one-dimensional structures, built on a two-dimensional
semiconductor heterojunction surface and with the carrier density
modulation by an applied gate potential.

Generally there are two standard types of one-dimensional
conductive structure: a quantum-point-contact system and a ballistic
one-dimensional quantum wire made by the cleaved-edge overgrowth technique
\cite{Z}.
Both these structures show pronounced quantized steps of conductance
when the gate potential modulates the carrier density in the channel. 

We remarked above that a one-dimensional electronic system will show
quantized conductance plateaus corresponding to the accessing of
distinct sub-bands in the channel. Once the
chemical potential moves up with the applied gate voltage, at some value
the next upper sub-band starts to fill. The explanation
of this basic physical effect is clear within various
models.
The literature already contains an abundance of
observations and their explanation.

Beyond this basic outcome several finer-scale anomalies appear.
One is the well observed zero bias anomaly. A tunnelling picture
at nearly zero source-drain bias will reproduce this
anomaly. In this case explanations by quasi-particle tunnelling
and topological Majorana states have brought some clear understanding,
but have given cause for some theoretical controversy.

Matters are quite different for the main issue we have discussed,
concerning the so-called 0.7 anomaly in the value of the
conductance at some threshold in the channel chemical potential.
Despite many claims to universality, it is clear that the anomaly at
the non-integral conductance value is observed to be non-universal
and non-generic. It shows itself as material-sensitive and also
sensitive to temperature and applied magnetic field.

We have discussed the received reasons for this anomaly, as reflected
in its considerable literature. Of the leading models, the
spin-polarised band picture requires a pure 1D electron gas
while, by contrast, the alternative Kondo picture needs the
existence of a localised impurity state. Although these two explanations
are mutually exclusive, published measurements can be adduced
in support of both. Therefore there is no consensus, either
theoretically or even in terms of any arguable ``universality''
in observations of the 0.7 anomaly.

In our opinion the van-Hove scenario is an acceptable alternative
towards an account of the anomaly, if only because every one-dimensional
electronic system must exhibit -- universally -- singularities at the band
edges. Near band energies at the edge, the occurrence of
high densities of states will influence the conductance profile.
As a result an anomaly is possible at a plateau-to-plateau
crossover in carrier density. This is sometimes known as the ridge-state
anomaly. We foreshadow a detailed trial calculation based on this picture.

\section*{Acknowledgment}

MPD thanks Professor Nguyen Van Hieu and the Vietnam Academy
of Science and Technology for their kind hospitality during
the author's participation in the Eighth International Workshop
on Advanced Materials Science and Nanotechnology, Ha Long City, Vietnam.

\end{document}